\documentclass[aps,prd,reprint]{revtex4-2}

\usepackage{xspace}
\usepackage{lipsum}
\usepackage{relsize}
\usepackage{appendix}
\usepackage{amsmath,amstext,amssymb,amsfonts}
\usepackage{epsfig}
\usepackage{url}
\usepackage{nameref}
\usepackage{varioref}
\usepackage{hyperref}
\usepackage{cleveref}
\usepackage[normalem]{ulem} 

\usepackage{latexsym}
\usepackage{epsfig}
\usepackage{wasysym}
\usepackage{graphicx}
\usepackage{verbatim}
\usepackage{tikz} 
\usepackage{pgfplots}
\usepackage[export]{adjustbox}

\usepackage{bbold}

\usepackage[normalem]{ulem}

\newcommand{\CMU}{McWilliams Center for Cosmology, Department of Physics, Carnegie Mellon University, Pittsburgh, PA 15213, USA}

\graphicspath{{./}{figures/}}

\begin{document}

\title{Spectral siren cosmology from gravitational-wave observations in GWTC-4.0}

\author{Ignacio Maga\~na~Hernandez}
\email{imhernan@andrew.cmu.edu}
\affiliation{\CMU}

\author{Antonella Palmese}
\email{apalmese@andrew.cmu.edu}
\affiliation{\CMU}

\begin{abstract}
Gravitational wave standard sirens offer a promising avenue for cosmological inference, particularly in measuring the expansion history of the universe. Traditionally, bright sirens require an electromagnetic counterpart to determine the redshift of the emission source while dark sirens rely on the presence of complete galaxy catalogs over large sky regions. Spectral sirens, using GW data alone, can circumvent these limitations by leveraging features in the mass distribution of compact binaries. With the recent release of the Gravitational-Wave Transient Catalog 4 (GWTC-4.0), the number of significant binary black hole (BBH) merger candidates has increased to 153, enabling more robust population studies and cosmological constraints. This work builds upon previous spectral siren analyses by analyzing the latest BBH observations with parametric and non-parametric models. In particular, we consider a parametric approach using the \textsc{Powerlaw + Peak} and \textsc{Broken Powerlaw + 2 Peaks} models as well as a more flexible non-parametric model based on Gaussian processes. We find broad consistency in the inferred Hubble constant $H_0$ constraints across models. Our most constraining result is from the \textsc{Gaussian Process} model, which, combined with the GW170817 bright siren measurement, results in $H_0 = 69^{+7}_{-6} \ \mathrm{km\,s^{-1}\,Mpc^{-1}}$, a 10\% precision measurement. For the \textsc{Powerlaw + Peak} and \textsc{Broken Powerlaw + 2 Peaks} we find fractional uncertainties of 17\% and 13\% respectively.

\end{abstract}

\date{\today}
\maketitle

\section{Introduction}~\label{sec:Intro}
The recent release of the Gravitational-Wave Transient Catalog 4.0 (GWTC-4.0;~\cite{GWTC4-catalog}) by the LIGO/Virgo/KAGRA (LVK) Collaboration has expanded the number of gravitational-wave (GW) candidates to a total of 218 with a probability of astrophysical origin $p_{\mathrm{astro}} \geq 0.5$. Among these, 153 are confident binary black hole (BBH) mergers with false alarm rates (FAR) less than one per year~\cite{GWTC4-catalog, GWTC4-methods}.

The latest population analyses from the LVK Collaboration have examined the properties of merging compact binaries using the GWTC-4.0 dataset~\cite{GWTC4-pop}. In particular, the binary black hole population reveals a feature-rich mass spectrum, with statistically significant deviations from simple power-law models. For example, previous analyses using GWTC-3~\cite{GWTC3-catalog} observations identified a robust excess near $\sim35\,M_\odot$ in the primary mass distribution~\cite{GWTC2-pop,GWTC3-pop}. However, the inclusion of new events in GWTC-4.0 reveals the presence of additional structure~\cite{GWTC4-pop}, motivating the adoption of more flexible parametric~\cite{Talbot:2018cva,Farah:2023swu,Golomb:2023vxm,MaganaHernandez:bumps,Roy:2025ktr} and non-parametric~\cite{Edelman:2022ydv,Tiwari:2021yvr,Ray:2023upk,Ray:2024hos,Callister:2023tgi,Heinzel:2024jlc,Antonini:2025zzw,maganaHernandez:massgap} models to capture the underlying astrophysical population. 

Gravitational-wave observations are standard sirens~\cite{Schutz:1986gp,Holz:2005df}, as these provide direct measurements of luminosity distance and redshifted detector-frame masses. The detector-frame masses and luminosity distance can be related to the source-frame masses and redshift, by assuming a cosmological model. Astrophysically motivated models for the source-frame mass distribution of black holes, particularly those featuring distinct structures such as mass gaps or bumps, can be used to statistically infer the redshift of the source and thereby breaking the mass-redshift degeneracy~\cite{Taylor:2012db,Farr:2019twy}. This approach underpins the spectral siren method~\cite{Ezquiaga:2022zkx}, which leverages population-level features in the mass spectrum to constrain cosmology and modified GW propagation effects using gravitational-wave data alone~\cite{Ezquiaga:2021ayr,MaganaHernandez:2021zyc,Leyde_2022,gwtc3-cosmo}. For reviews on GW cosmology, including the spectral siren method we refer the reader to Refs~\cite{Palmese:2025zku,Chen:2024gdn}.

The accuracy of this method depends on the fidelity of the inferred mass distribution. Missed or misidentified features can lead to biased cosmological inference~\cite{Pierra:2023deu}. To mitigate this, flexible and model-independent population inference techniques, such as non-parametric Gaussian Process (GP) models or hybrid parametric models with localized perturbations, offer a robust alternative to more rigid parametric models~\cite{Tiwari:2021yvr,Edelman:2022ydv,Ray:2023upk,Callister:2023tgi,Ray:2024hos,Heinzel:2024jlc,Farah:2024xub,Antonini:2025zzw,maganaHernandez:massgap}. Notably, the binned Gaussian process (BGP) model used in Refs.~\cite{Ray:2023upk,Ray:2024hos} has enabled the first data-driven spectral siren cosmological measurement using GWTC-3 BBH observations~\cite{MaganaHernandez:2024uty}, yielding competitive constraints on the Hubble constant when compared to using parametric models.

In this paper, we perform the first spectral siren cosmological inference using the latest gravitational-wave observations from the GWTC-4.0 data release. We apply three distinct models to characterize the BBH mass distribution: The previously preferred LVK model \textsc{Powerlaw + Peak} in GWTC-3~\cite{GWTC3-pop}, the \textsc{Broken Powerlaw + 2 Peaks} model currently preferred in the LVK GWTC-4.0 population analysis~\cite{GWTC4-pop}, as well as a non-parametric astrophysics informed GP (here after denoted as \textsc{Gaussian Process}) model~\cite{maganaHernandez:massgap}. We compare our results to state of the art measurements of cosmological parameters from the Planck~\cite{aghanim2020planck} and SH0ES~\cite{riess2022comprehensive} collaborations. We provide joint constraints on the Hubble constant when combined with its measurement using the bright siren GW170817~\cite{TheLIGOScientific:2017qsa,abbott2017gravitational,LIGOScientific:2017adf,abbott2019properties,Palmese:2023beh} and its host galaxy NGC 4993. We discuss potential sources of systematics and provide a discussion for future directions.

\section{Methods}~\label{sec:Methods}
In this section, we describe in detail the spectral siren framework used to perform cosmological inference using features in the mass spectrum of the binary black hole population. We also outline its connection to hierarchical Bayesian inference (for a review see Ref.~\cite{Vitale:2020aaz}), which is employed to jointly constrain both the BBH population and the underlying cosmological parameters. Additionally, we present the mass and redshift models adopted in this work.

\subsection{Spectral Siren Framework}
Gravitational-wave observations of BBH mergers provide direct measurements of the redshifted, detector-frame component masses, $m_1^{\text{det}}$ and $m_2^{\text{det}}$, as well as the luminosity distance to the emission source, $D_L$. To infer the corresponding source-frame masses $m_1$ and $m_2$, one must assume a cosmological model characterized by a set of cosmological parameters $\Lambda_c$, which allows for the computation of the redshift $z$ as a function of luminosity distance, $z(D_L|\Lambda_c)$. The source-frame masses are given by
\begin{equation}
    m_{1,2} = \frac{m_{1,2}^{\text{det}}}{1 + z(D_L|\Lambda_c)},
\end{equation}
such that different choices of cosmological parameters $\Lambda_c$ yield different estimates of the redshift and source-frame masses. By jointly modeling the BBH population and the cosmological parameters, we can self-consistently infer the mass and redshift distribution of BBHs and constrain cosmology using GW data alone. This approach forms the basis of spectral siren cosmology, which leverages population-level features to break the degeneracy between mass and redshift~\cite{Taylor:2012db,Farr:2019twy,Ezquiaga:2022zkx}.

We can express the number density of BBH mergers in terms of their detector-frame parameters, denoted by ${\theta}^{\text{det}}$, and relate it to the corresponding source-frame parameters, ${\theta}^s$, via a change of variables. Specifically, the differential number density in detector-frame coordinates is given by

\begin{equation}
    \frac{dN({\theta}^{\text{det}} | \Lambda_p, \Lambda_c)}{d{\theta}^{\text{det}}} = \frac{dN({\theta}^s | \Lambda_p)}{d{\theta}^s} \left| \frac{d{\theta}^s}{d{\theta}^{\text{det}}}(\theta^{\text{det}}|\Lambda_c) \right|.
\end{equation}

\noindent
Here, $\Lambda = \{\Lambda_p, \Lambda_c \}$ denotes the set of population and cosmological parameters, respectively. The term $\left| {d{\theta}^s}/{d{\theta}^{\text{det}}} \right|$ is the Jacobian determinant of the transformation from detector-frame to source-frame variables. This Jacobian encodes the dependence of the source-frame quantities on the cosmological model and is essential for correctly mapping the observed distribution to the underlying astrophysical population. Therefore, spectral siren cosmology is a simple extension of well-established hierarchical Bayesian population inference in the source frame, where the underlying cosmological model is varied at the inference level.

\subsection{Hierarchical Population Inference}
\label{sec:hierarchical}

Given the transformation described above, we work with detector-frame parameters $\theta^{\text{det}} = \{m_1^{\text{det}}, m_2^{\text{det}}, D_L\}$ and compute the corresponding source-frame parameters $\theta^s = \{m_1, m_2, z\}$ under a cosmological model with parameters $\Lambda_c$. This enables a self-consistent fit for the BBH population distribution described by a model with parameters $\Lambda_p$. We adopt the usual hierarchical Bayesian framework for population inference, following the methodology outlined in~\cite{Thrane_2019, Vitale:2020aaz}, and extended for spectral siren cosmology in~\cite{Taylor:2012db,Farr:2019twy}.

The differential number density of BBH mergers in the source-frame is
\begin{equation}
    \frac{dN(m_1, m_2, z | \Lambda)}{dm_1\,dm_2\,dz} \propto \frac{dV_c}{dz} \left( \frac{T_{\mathrm{obs}}}{1+z} \right) \mathcal{R}_0 \psi(z|\Lambda)p(m_1, m_2 | \Lambda),
\end{equation}
\noindent
where we now use $\Lambda$ for simplicity. The term $dV_c/dz$ is the differential comoving volume, $T_{\mathrm{obs}}$ is the total observation time and the factor $1/(1+z)$ converts source-frame time to detector-frame time. The local merger rate (at $z=0$), $\mathcal{R}_0$, is marginalized and evolves with redshift according to $\psi(z|\Lambda) = (1 + z)^\kappa$ where $\kappa$ is its powerlaw slope~\cite{Fishbach:2018edt}. The mass distribution $p(m_1, m_2 | \Lambda)$ is defined by the population model described in Section~\ref{sec:models}.

The posterior on the hyperparameters $\Lambda$ given the observed data $\{d_i\}$ from $N_{\mathrm{obs}}$ events is
\begin{align}
\begin{split}
    p(\Lambda | \{d_i\}) &\propto \frac{p(\Lambda)}{\beta(\Lambda)^{N_{\mathrm{obs}}}} \prod_{i=1}^{N_{\mathrm{obs}}} \left[ \int \mathcal{L}(d_i | m_1, m_2, z)\, \right. \\
    &\left. \times \frac{dN(m_1, m_2, z | \Lambda)}{dm_1\,dm_2\,dz}\, dm_1\,dm_2\,dz \right],
\end{split}
\end{align}
\noindent
where $\mathcal{L}(d_i | m_1, m_2, z)$ is the likelihood of the data $d_i$ given the source-frame parameters, and $\beta(\Lambda)$ accounts for selection effects by quantifying the fraction of detectable sources under the model with parameters $\Lambda$.

The integrals over individual event likelihoods are approximated using importance sampling from posterior samples obtained via standard parameter estimation pipelines. To compute $\beta(\Lambda)$, we use injection campaigns based on LVK sensitivity estimates~\cite{Tiwari_2018, Farr_2019, essick2022precisionrequirementsmontecarlo}, ensuring convergence through importance sampling diagnostics.

\subsection{BBH Mass Model}
\label{sec:models}

We model the joint BBH mass distribution over primary and secondary masses as,
\begin{equation}
    p(m_1, m_2) \propto p_{\text{BH}}(m_1)\, p_{\text{BH}}(m_2)\, f(m_1, m_2).
\end{equation}
\noindent
Here, $p_{\text{BH}}$ represents the common mass distribution of individual black holes, and $f$ encodes correlations between the component masses, a pairing function~\cite{Fishbach:2019bbm,Farah:2023swu}. We make the choice of pairing function,
\begin{equation}
    f(m_1, m_2) = \left( \frac{m_2}{m_1}\right)^{\beta}
\end{equation}
to depend on the mass ratio $q = m_2/m_1$ with powerlaw slope $\beta$.

For the common BH mass distribution $p_{\text{BH}}$, we consider the following models: (1) The previously preferred \textsc{Powerlaw + Peak} model from GWTC-3~\cite{Talbot:2018cva,GWTC3-pop}, (2) the currently preferred \textsc{Broken Powerlaw + 2 Peaks} in GWTC-4.0~\cite{GWTC4-pop} and the astrophysics informed \textsc{Gaussian Process}) model from~\cite{maganaHernandez:massgap}. For explicit details on the functional forms forms for each of these models we refer the reader to the referenced works. 

\subsection{Cosmological Model}
\label{sec:cosmo}

We adopt a flat $\Lambda$CDM cosmological model, characterized by a set of parameters $\Lambda_c = \{H_0, \Omega_m\}$. The redshift $z$ is related to the luminosity distance $D_L$ through the standard cosmological relation
\begin{equation}
    D_L(z | \Lambda_c) = (1 + z) \int_0^z \frac{c\, dz'}{H(z' | \Lambda_c)},
\end{equation}
\noindent
where $H(z | \Lambda_c) = H_0 \sqrt{\Omega_m(1+z)^3 + (1-\Omega_m)}$ is the Hubble parameter as a function of redshift, and $c$ is the speed of light. 

\section{Results}~\label{sec:Results}
We analyze 152 binary black hole detections reported in GWTC-4.0~\cite{GWTC4-catalog} that have false alarm rates less than one per year. We exclude GW231123 ~\cite{GW231123} from our sample for two reasons, it is the only event with masses above $100~ M_\odot$ making a population fit in this region uninformative and also due to its single event parameter uncertainty arising from waveform systematics~\cite{GW231123}. We have verified that inclusion of this event does not significantly change the results of this work.
These events constitute a confident BBH sample and provide a robust dataset for spectral siren cosmology~\cite{GWTC4-pop}. We use the injections released with GWTC-4.0 to estimate the detectable fraction and take selection effects into account~\cite{Essick:2025zed}. We apply the hierarchical Bayesian inference framework described in Section~\ref{sec:hierarchical}, using the BBH population models introduced in Section~\ref{sec:models} and the cosmological model outlined in Section~\ref{sec:cosmo}. This analysis allows us to jointly constrain the source-frame mass and redshift BBH distributions as well as cosmological parameters using gravitational-wave data alone. Throughout this work we use uniform priors on $H_0$ in the range $[20,\,220]\,\mathrm{km\,s^{-1}\,Mpc^{-1}}$
and on $\Omega_m$ in the range $[0,1]$. We report 68\% credible intervals (CI) with maximum a-posteriori values throughout (unless stated otherwise).

\begin{figure*}[ht]
    \centering
    \begin{minipage}[t]{0.49\textwidth}
        \centering
        \includegraphics[width=\linewidth]{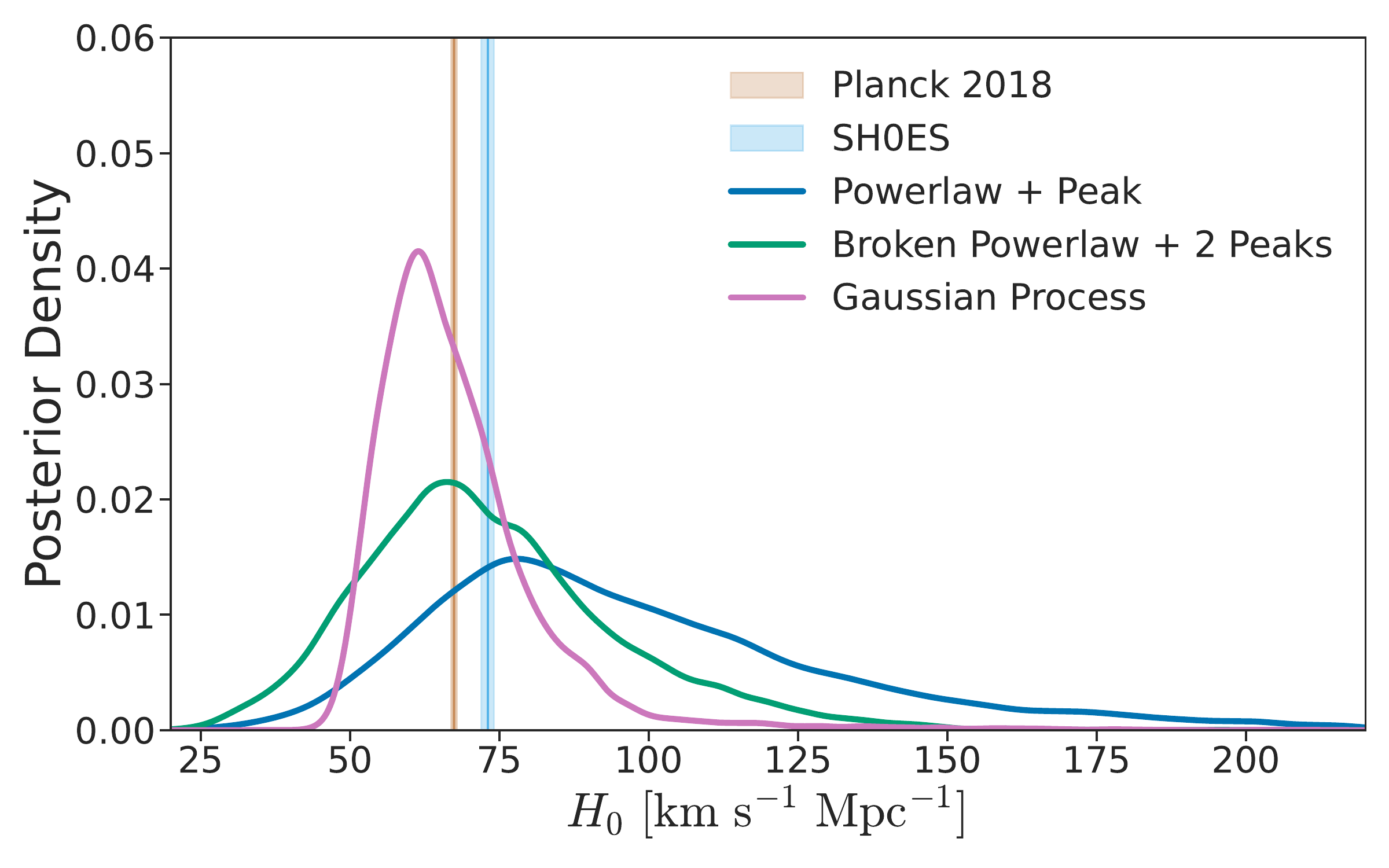}
    \end{minipage}
    \hfill
    \begin{minipage}[t]{0.49\textwidth}
        \centering
        \includegraphics[width=\linewidth]{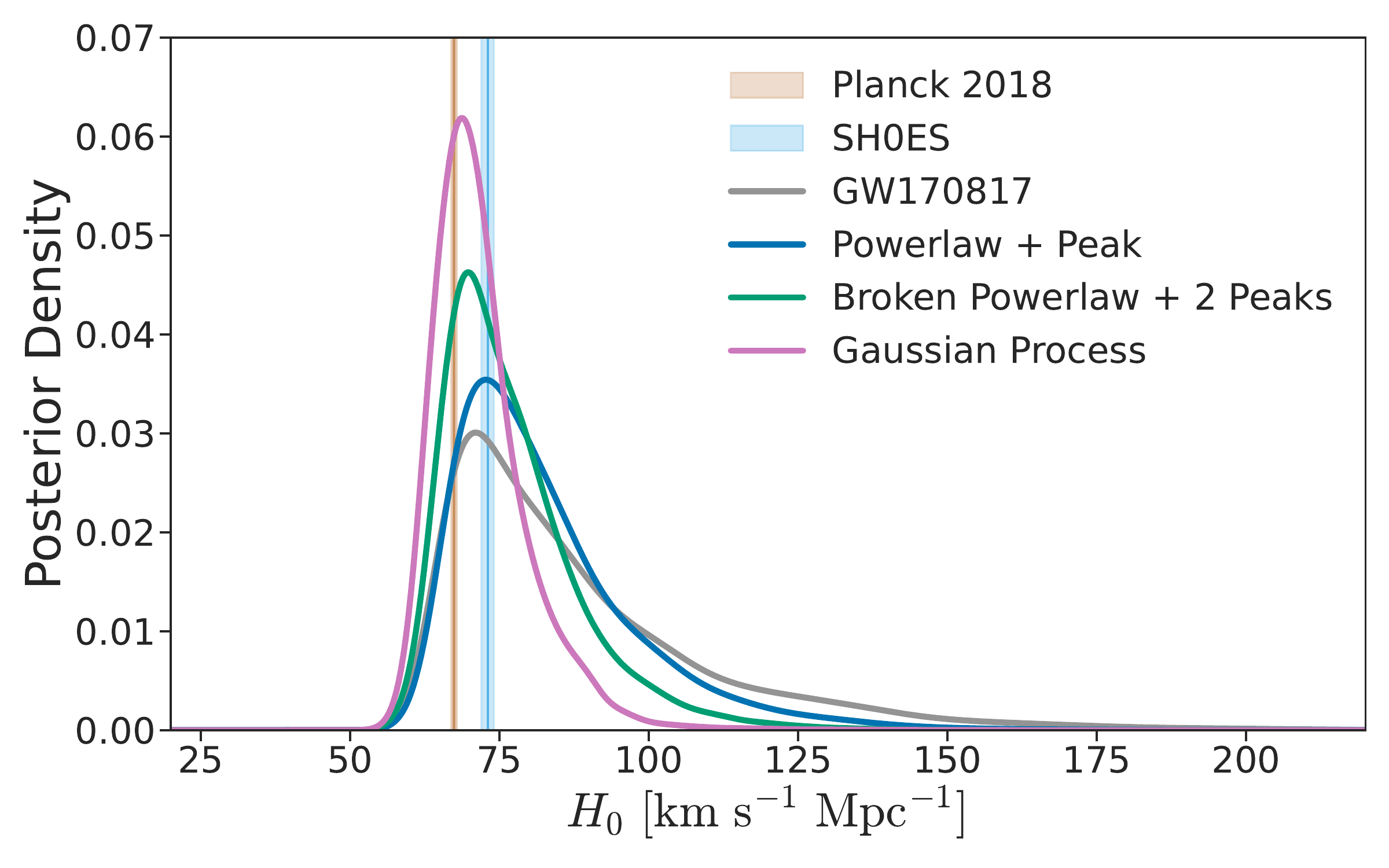}
    \end{minipage}
    \caption{Posterior distributions on the Hubble constant $H_0$ inferred using three different BBH population models: \textsc{Powerlaw + Peak} (Blue), \textsc{Broken Powerlaw + 2 Peaks} (Green), and \textsc{Gaussian Process} (Purple). The left panel shows constraints obtained using BBH data alone, while the right panel incorporates additional information from the binary neutron star merger GW170817 as a bright standard siren. The orange band shows percent level measurements on $H_0$ from the CMB~\cite{aghanim2020planck} and in the light blue shaded band from standard candle type 1A SN measurements~\cite{riess2022comprehensive}}
    \label{fig:h0_posteriors}
\end{figure*}

\begin{table}[ht]
\centering
\begin{tabular}{lccc}
\hline
\textbf{Model} & \textbf{$H_0$ (BBH)} & \textbf{$H_0$} & \textbf{$\sigma_{H_0}/H_0$} \\
\hline
\textsc{Powerlaw + Peak} & $78^{+93}_{-33}$ & $73^{+16}_{-8}$ & $17\%$ \\
\textsc{Broken Powerlaw + 2 Peaks} & $66^{+48}_{-32}$ & $70^{+12}_{-6}$ & $13\%$ \\
\textsc{Gaussian Process} & $61^{+29}_{-13}$ & $69^{+7}_{-6}$ & $10\%$ \\
\hline
\end{tabular}
\caption{Hubble constant estimates (in units of $\mathrm{km\,s^{-1}\,Mpc^{-1}}$) for different population models using only BBH observations. The second $H_0$ column provides our constraints combined with the bright siren estimate from GW170817, along with its estimated fractional uncertainty $\sigma_{H_0}/H_0$.}
\label{tab:results}
\end{table}

In Figure~\ref{fig:h0_posteriors}, we present the posterior distributions on the Hubble constant $H_0$ inferred using the three BBH population models described in Section~\ref{sec:models}. We show the posteriors on $H_0$ using the BBH observations with GWTC-4.0 alone as well as the joint posteriors when these are combined with the $H_0$ measurement from bright siren GW170817 and its host galaxy NGC 4993~\cite{TheLIGOScientific:2017qsa}. The GW170817 posterior we use corresponds to a measurement of $H_0 = 71^{+23}_{-8} \ \mathrm{km\,s^{-1}\,Mpc^{-1}}$ obtained from Ref.~\cite{,LIGOScientific:2017adf}. We also compute the fractional uncertainty for our each of our combined $H_0$ estimates. All of our results are summarized in Table~\ref{tab:results}. 

\begin{figure}[ht]
    \centering
    \includegraphics[width=\linewidth]{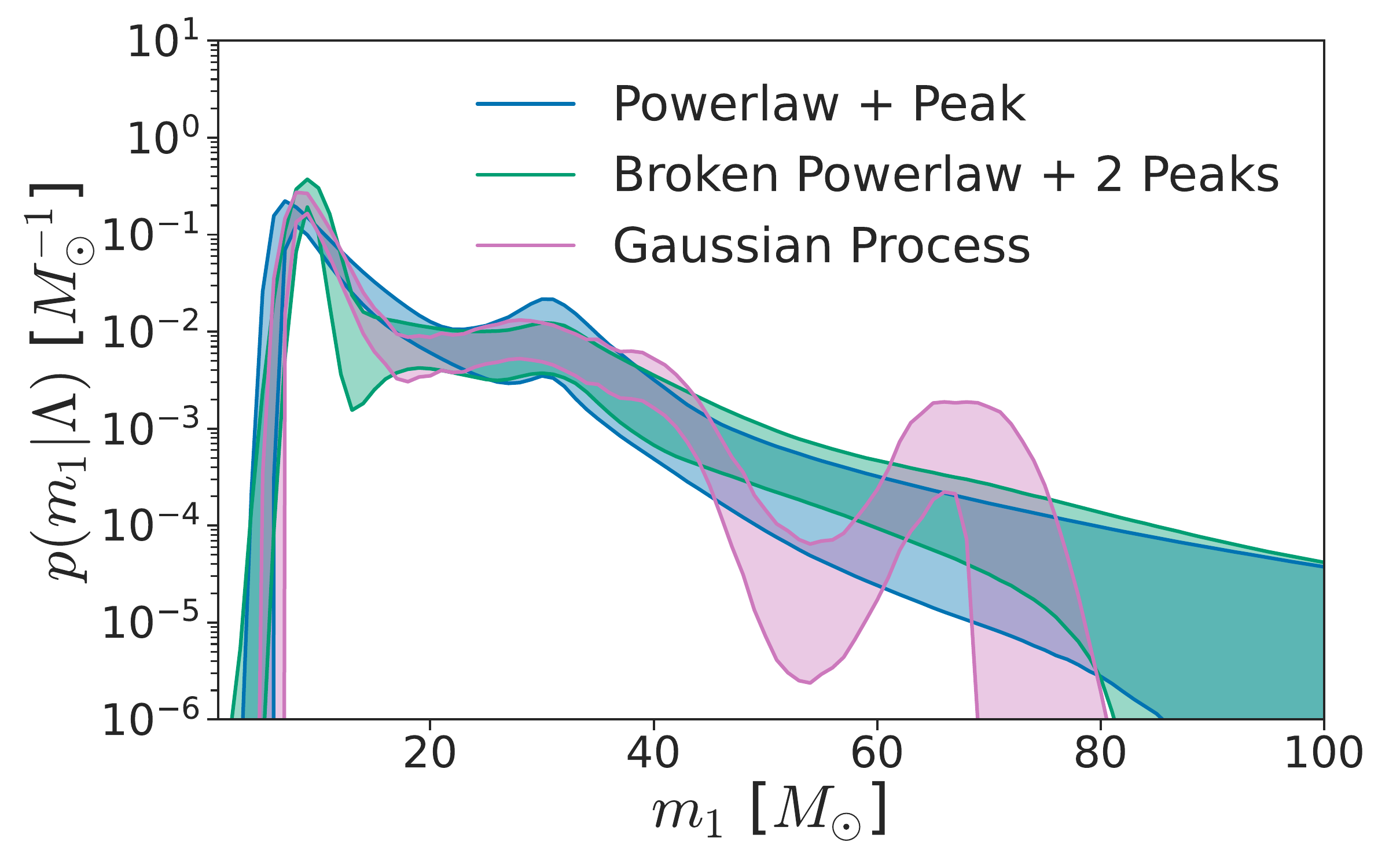}
    \includegraphics[width=\linewidth]{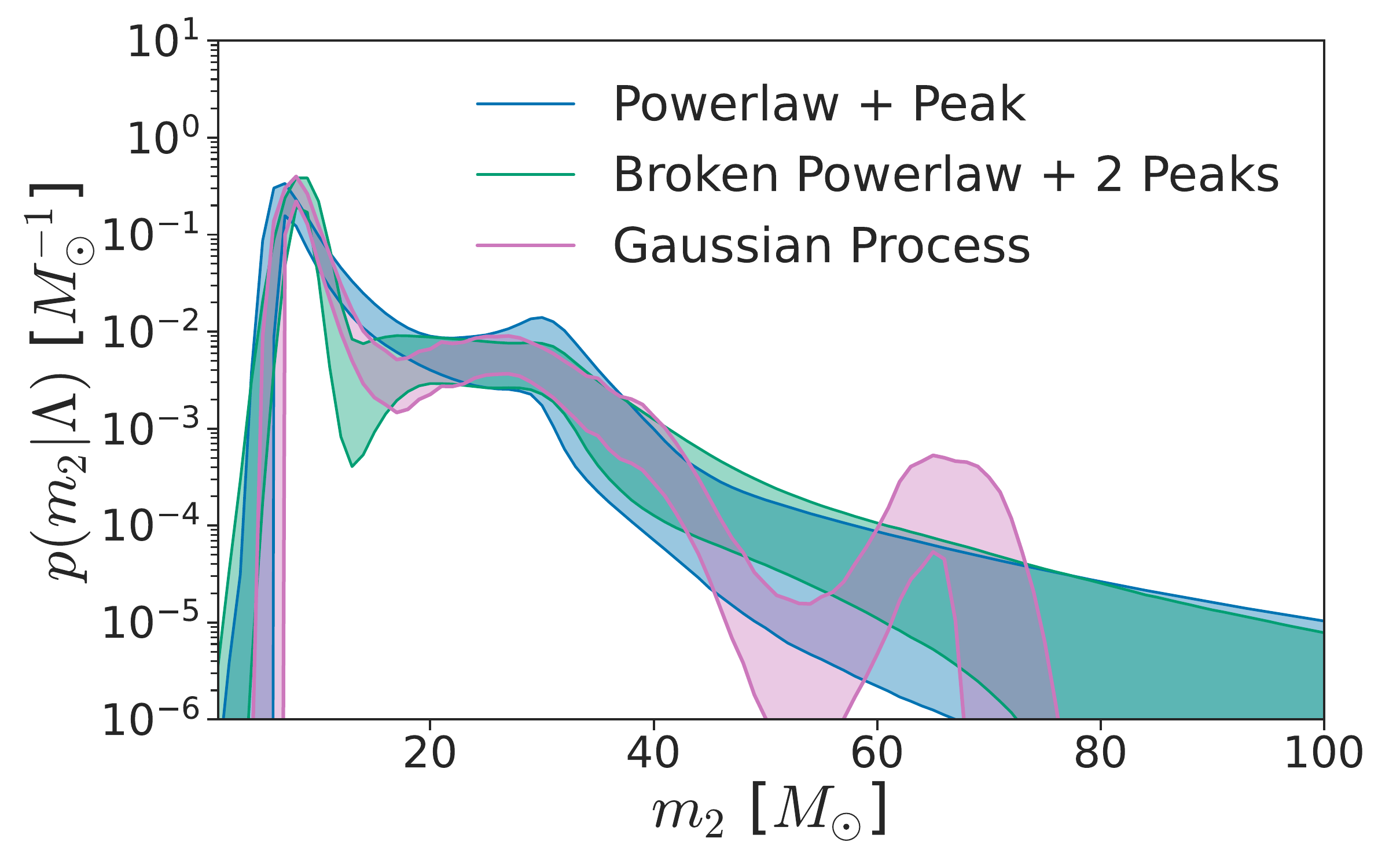}
    \includegraphics[width=\linewidth]{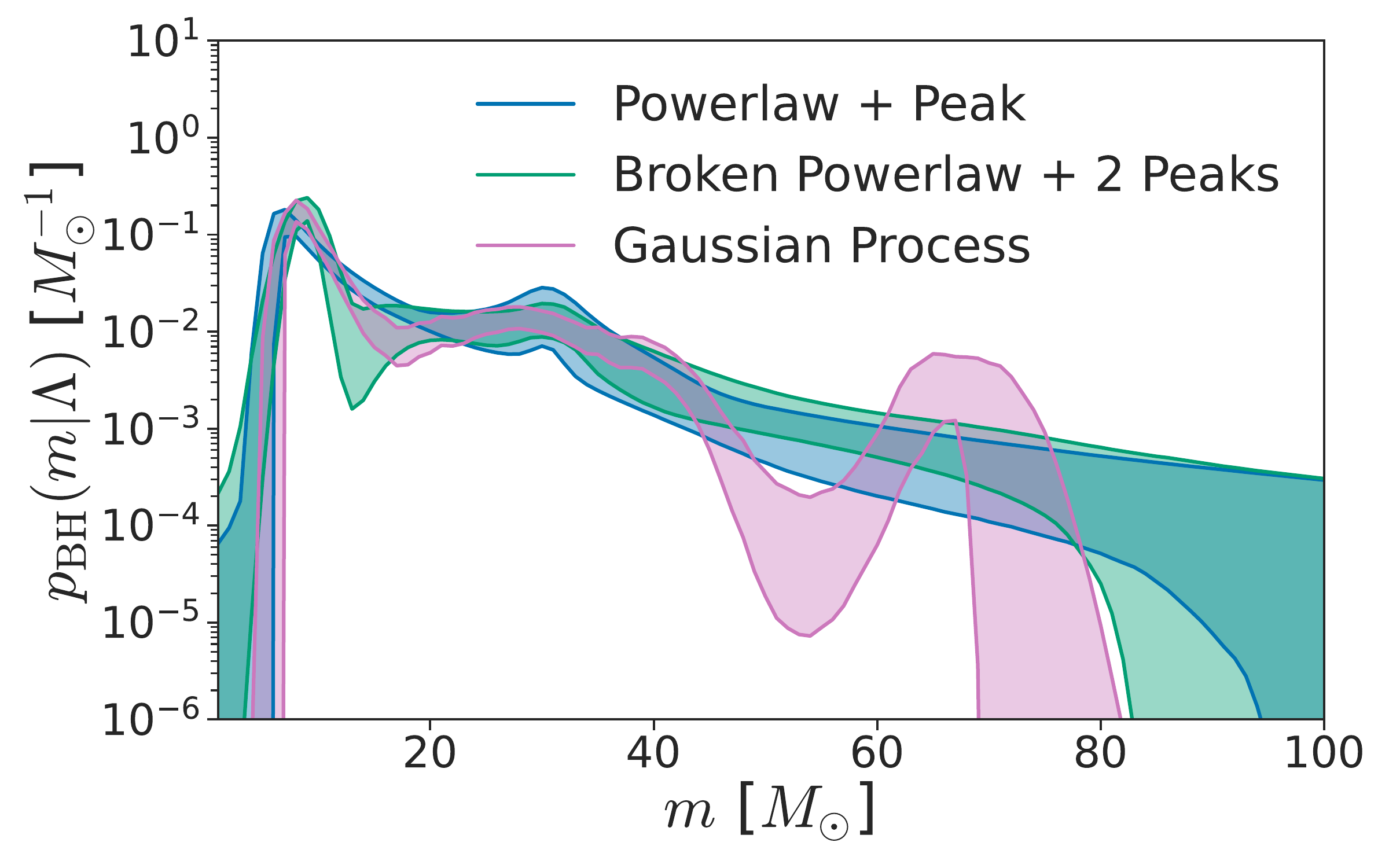}

    \caption{Inferred primary (top panel), secondary (middle panel) marginalized black hole mass distributions as well as the common black hole mass function (bottom panel) for each of the three population models considered in this work: \textsc{Powerlaw + Peak} (Blue), \textsc{Broken Powerlaw + 2 Peaks} (Green), and \textsc{Gaussian Process} (Purple). Shaded regions show 95\% credible intervals in the inferred shape of each mass distribution. }
    \label{fig:mass}
\end{figure}

In Figure~\ref{fig:mass}, we show the inferred primary and secondary (marginalized) black hole mass distributions, along with the underlying astrophysical BH mass function for each of our models. Distinct features emerge across the models analyzed in this work. For the \textsc{Powerlaw + Peak} model, we recover the expected peak near $35\,M_\odot$, which primarily drives the spectral siren $H_0$ inference~\cite{gwtc3-cosmo,Mali:2024wpq}, with additional contributions from the minimum BH mass measurements. In the \textsc{Broken Powerlaw + 2 Peaks} model, we observe more complex structure in the $10$--$40\,M_\odot$ range, consistent with population analyses from the LVK~\cite{GWTC4-pop}. This additional structure, in particular, the $\sim\,10 M_\odot$ peak shifting to higher masses, leads to lower inferred values of $H_0$ when compared to the results using the \textsc{Powerlaw+Peak} model. The $\sim35\,M_\odot$ peak is still present, albeit with reduced significance compared to Ref.~\cite{GWTC3-pop}. Both of these models do not allow for suppression of the BBH merger rate above $\sim45\,M_\odot$, as the power-law components above this range continue uninterrupted, other than allowing for a maximum BH mass which we find to be  $m_{\text{max}} > 80\,M_\odot$, consistent with the events in this mass range.

For the \textsc{Gaussian Process} model, we observe similar structure in the $10$--$40\,M_\odot$ range as seen in the \textsc{Broken Powerlaw + 2 Peaks} model. Under this mode, the $\sim 10\,M_\odot$ peak shifts to higher masses, which also explains the lower $H_0$ values we infer, consistent with the \textsc{Broken Powerlaw + 2 Peaks} results. However, the flexibility of the \textsc{Gaussian Process} model also captures a shoulder-like feature around $\sim45\,M_\odot$ with a suppression of the BBH merger rate thereafter. These features are consistent with the analysis in Ref.~\cite{maganaHernandez:massgap}, which associates them with the onset of the PISN mass gap and the pulsational PISN buildup expected at the lower mass gap edge. We also observe a population of high mass mergers in the $\sim60-70\,M_\odot$ range, likely a contribution due to hierarchical mergers populating a region of the PISN mass gap, consistent with previous analyses~\cite{MaganaHernandez:bumps,Antonini:2025zzw,GWTC4-pop,maganaHernandez:massgap}. These features lead to a more informative $H_0$ constraint compared to the measurements done with the parametric models discussed in this work. Future work should further explore these features using GWTC-4.0 data and the \textsc{Gaussian Process} model, extending the results of Ref.~\cite{maganaHernandez:massgap} under a fixed cosmological model.

\section{Discussion and Conclusions}~\label{sec:Conclusions}
In this paper, we analyzed the latest GWTC-4.0 BBH observations to present the first spectral siren measurement using this dataset. We employ the preferred parametric models from GWTC-3, the \textsc{Powerlaw + Peak} model, and from GWTC-4.0, the \textsc{Broken Powerlaw + 2 Peaks} model, to perform joint population and cosmological inference. Additionally, we incorporate a flexible non-parametric approach using the \textsc{Gaussian Process} model.

The increase in complexity and flexibility of the models presented in this work leads to tighter constraints on $H_0$ when combined with the bright standard siren $H_0$ measurement from GW170817. As GWTC-4.0 has shown, the \textsc{Powerlaw + Peak} model does not provide a good fit to the observed data~\cite{GWTC4-pop}, it is not surprising that the BBH-only spectral siren constraint using this model yields less informative (and potentially biased) $H_0$ estimates compared to the \textsc{Broken Powerlaw + 2 Peaks} and \textsc{Gaussian Process} models, which show a slight preference for lower $H_0$ values. Simulated studies, such as Ref.~\cite{Pierra:2023deu}, have demonstrated that model misspecification can lead to such biases.

The latest observations from GWTC-4.0 have revealed a more complex population distribution for BBHs. The mass spectrum exhibits an intricate structure, for example, the $\sim 35~M_\odot$ bump is not well modeled by a simple Gaussian excess~\cite{GWTC4-pop}. Additionally, evidence for the pair-instability supernova mass gap~\cite{maganaHernandez:massgap,Antonini:2025zzw,GW231123}, along with a subpopulation that appears to populate this gap~\cite{MaganaHernandez:bumps}, introduces distinct features in the mass spectrum. These features serve as characteristic mass scales that can tighten constraints from spectral sirens, as can be seen from the results using the \textsc{Gaussian Process} model. To quantify their contribution, the framework proposed in Ref.~\cite{Mali:2024wpq} can be useful for assessing how much information each feature provides for cosmological parameter inference.

A caveat in our analysis is the assumption that the BBH mass distribution does not evolve with redshift, consistent with previous works~\cite{gwtc3-cosmo,MaganaHernandez:2024uty}. This assumption is motivated by the latest GWTC-4.0 population constraints, which show no evidence of redshift evolution in the mass spectrum when using the BGP model~\cite{GWTC4-pop,Ray:2023upk}. However, if certain spectral siren features in the mass spectrum do evolve with redshift, studies such as Refs.~\cite{Ezquiaga:2022zkx,agarwal2025blindedmockdatachallenge,Tong:2025xvd} have demonstrated that this can introduce biases in the inference of cosmological parameters. To date, no spectral siren analysis has successfully produced estimates of $H_0$ using non-parametric methods that account for potential redshift evolution in the BH mass distribution. To address this, future work should extend the \textsc{Gaussian Process} model to enable joint inference of mass and redshift. 

While the precision of current spectral siren measurements is getting close to those from dark siren measurements using the aid of galaxy catalogs \cite{Palmese:2021mjm,gwtc3-cosmo,Bom:2024afj}, future observations using current and next generation GW detectors, have the potential to constrain the Hubble constant at the percent level using spectral sirens as a cosmological probe~\cite{Chen:2024gdn}. 

\section*{Acknowledgements}
The authors would like to thank Ariel Amsellem and Yeajin Kim for useful comments. IMH is supported by a McWilliams postdoctoral fellowship at Carnegie Mellon University. This material is based upon work supported by the National Aeronautics and Space Administration under Grant No. 22-LPS22-0025. This research has made use of data or software obtained from the Gravitational Wave Open Science Center (gwosc.org), a service of the LIGO Scientific Collaboration, the Virgo Collaboration, and KAGRA. This material is based upon work supported by NSF's LIGO Laboratory which is a major facility fully funded by the National Science Foundation, as well as the Science and Technology Facilities Council (STFC) of the United Kingdom, the Max-Planck-Society (MPS), and the State of Niedersachsen/Germany for support of the construction of Advanced LIGO and construction and operation of the GEO600 detector. Additional support for Advanced LIGO was provided by the Australian Research Council. Virgo is funded, through the European Gravitational Observatory (EGO), by the French Centre National de Recherche Scientifique (CNRS), the Italian Istituto Nazionale di Fisica Nucleare (INFN) and the Dutch Nikhef, with contributions by institutions from Belgium, Germany, Greece, Hungary, Ireland, Japan, Monaco, Poland, Portugal, Spain. KAGRA is supported by Ministry of Education, Culture, Sports, Science and Technology (MEXT), Japan Society for the Promotion of Science (JSPS) in Japan; National Research Foundation (NRF) and Ministry of Science and ICT (MSIT) in Korea; Academia Sinica (AS) and National Science and Technology Council (NSTC) in Taiwan.

\bibliography{references}{}
\bibliographystyle{aasjournal}

\end{document}